# Stellar Intensity Interferometric Capabilities of IACT Arrays

**Dave Kieda[1]**
*Department of Physics and Astronomy*
*University of Utah, Salt Lake City, Utah, USA*
*E-mail:* `dave.kieda@utah.edu`

**Nolan Matthews**
*Department of Physics and Astronomy*
*University of Utah, Salt Lake City, Utah, USA*
*E-mail:* `nolankmatthews@gmail.com`

Sub-milliarcsecond imaging of nearby main sequence stars and binary systems can provide critical information on stellar phenomena such as rotational deformation, accretion effects, and the universality of starspot (sunspot) cycles. Achieving this level of resolution in optical wavelength bands (U/V) requires use of a sparse array of interferometric telescopes with kilometer scale baseline separations. Current ground based VHE gamma-ray observatories, such as VERITAS, HESS, and MAGIC, employ arrays of > 10 m diameter optical Imaging Atmospheric Cherenkov Telescopes (IACTs) with >80 m telescope separations, and are therefore well suited for sub-milliarcsecond astronomical imaging in the U/V bands using Hanbury Brown and Twiss (HBT) interferometry [1,2]. We describe the development of instrumentation for the augmentation of IACT arrays to perform Stellar Intensity Interferometric (SII) imaging. Laboratory tests are performed using pseudo-random and thermal (blackbody) light to demonstrate the ability of high speed (250 MHz) digitizing electronics to continuously record photon intensity over long periods (minutes to hours) and validate the use of offline software correlation to calculate the squared degree of coherence $|\gamma(r)|^2$. We then use $|\gamma(r)|^2$ as the interferometric observable to populate the Fourier reciporical image plane, and apply standard inversion techniques to recover the original 2-D source image. The commercial availability of inexpensive fiber-optic based sub-nanosecond multi-crate (White Rabbit[3]) synchronization timing enables the extension of SII to baselines greater than 10 km, theoretically allowing U/V band imaging with resolution <100 $\mu$ arc-seconds. This article provides a description of typical designs of practical SII instrumentation for the VERITAS IACT observatory array (Amado, Arizona) and the future CTA IACT Observatory (Canary Islands, Spain and Paranal, Chile).

*35th International Cosmic Ray Conference - ICRC2017*
*10-20 July, 2017*
*Bexco, Busan, Korea*

We gratefully acknowledge financial support but the US National Science Foundation, and thank members of the VERITAS & CAT collaborations for helpful comments. We also thank Stephan LeBohec for his review of this article, and his kind assistance and guidance in many aspects of the experimental results. This work was conducted in the context of the CTA Consortium.

---

[1]Speaker

1. SII and sub-milliarc-second Astronomical Resolution

The diffraction-limited angular resolution $\theta$ of an astronomical telescope with diameter D is given by the Rayleigh criterion: $\theta = 1.22 \lambda/D$ . Optical astronomical imaging using amplitude interferometry (AI) measures the degree of coherence of light $\gamma(r)$ (a.k.a visibility) between telescopes with baselines up to 100-300 m, approaching milliarcsecond (mas) angular resolution in JHK wavebands[4]. Sub-mas angular resolution of stars at shorter wavelengths (400 nm < $\lambda$ < 800 nm) is possible using SII). SII can measure the squared degree of coherence $|\gamma(r)|^2$ with a sparsely populated telescope arrays containing baselines extending to 1 km or more. Because it is now possible to synchronize telescope data recording to < 1 nsec over 80 km distances using commercially available fiber optic adapters[3], it becomes feasible to perform optical astronomical observations with sub-mas resolution. Existing IACT arrays, such as the VERITAS Observatory [5,6,7] can be easily instrumented with SII capabilities to provide U/V band observations, combined with the increased angular resolution. It is also possible to develop U/V band observations with angular resolution well below 100 μas for future km-baseline IACT arrays (i.e. Cherenkov Telescope Array, CTA) [8,9].

The relative insensitivity of $|\gamma(r)|^2$ to atmospheric turbulence [8] allows SII to observe in the U/V bands, where the spectral density is maximal for massive, hot stars (O and B). The shorter wavelengths in these optical bands provides increased the angular resolution, as well as maximal contrast for temperature gradients and surface features on these massive stars. The combination of higher angular resolution and shorter wavelength sensitivity allows for multiple new science investigations in stellar astrophysics. Potential science topics that may be explored with a U/V imaging capability with sub-mas resolution include: Joint AI and SII measurements of stellar envelopes as a function of temperature[10]; rapidly rotating stars [11,12], including highly deformed stars that may change evolutionary trajectory [13] and create strong temperature gradients in hot stars [14,15]; hot stars with circumstellar decretion disks such as Be and B[e] stars [16,17]; Wolf-Rayet Stars and nearby colliding winds [18,19]; blue supergiants [20,21,22]; direct imaging of the evolution of interacting binaries [23,24]; direct observation of starspots, and exploration of the conditions linked to the appearance and periodicity of starspots in main sequence stars [25].

2.Laboratory tests of SII

We have built a laboratory at the University of Utah to explore SII instrumentation using high speed streaming data acquisition and software correlation. The SII laboratory uses a bright light source combined with a pinhole spatial mask to simulate the images of stars or binary systems (Figure 1). The light source is either a thermal source (Hg arc lamp with 435.8 nm G filter, 10nm width) or a pseudo-random source (543 nm laser with rotating ground glass). After passing through the pinhole mask, light travels through a 3m long dark box to an optical table, where a non-polarizing beam splitter separates the primary beam into two secondary beams.

The light intensity of the secondary beams is measured by a pair of super-bialkali (>35% Q.E.) high speed photomultiplier tube (PMT) (Hamamatsu R10560-100-20 MOD)[5]. One PMT is fixed in position; the other PMT is located on a computer controlled linear actuator, allowing computer programmable PMT movements to perform automated linear scans of the spatial correlation. Optical polarizers may be inserted in front of each PMT, to select specific polarization angles. Each PMT receives high voltage (HV) power from a fiber optic controlled, battery powered HV supply. This HV supply design eliminates potential spurious noise correlations created by unintentional ground loops.

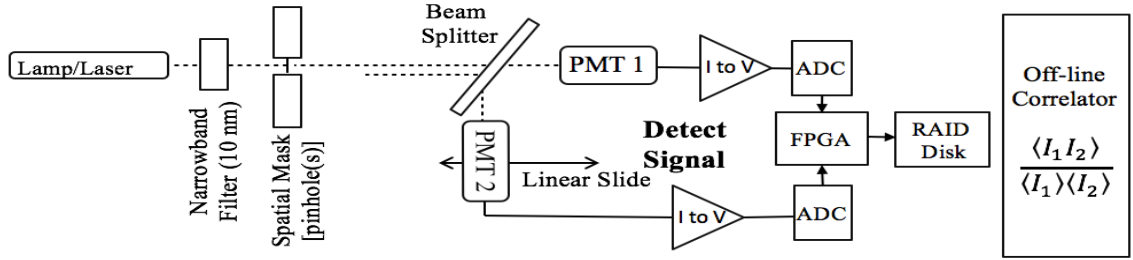

*Figure 1: Utah SII Test laboratory instrumentation. The 10nm filter is only used for the Hg Lamp source.*

The output from each PMT is fed into a high speed (>200 MHz) FEMTO preamplifier, and the resultant signal is continuously digitized (250 Mhz) using a National Instrument FlexRio Module (NI-5761). The digitized data from each PMT is scaled, truncated to 8 bits, and merged into a single continuous datastream by a Virtex-5 FGPA (PXIe-7965). The datastream is recorded to a high speed (700 MB/sec) 12 TB RAID disk. Data is typically recorded for 10 sec to 10 min durations, although the size of the RAID disk allows continuous recording for up to six hours. After the data recording is complete, software gain correction and noise reduction filtering are applied to each data channel, and cross-correlations and auto-correlations are calculated.

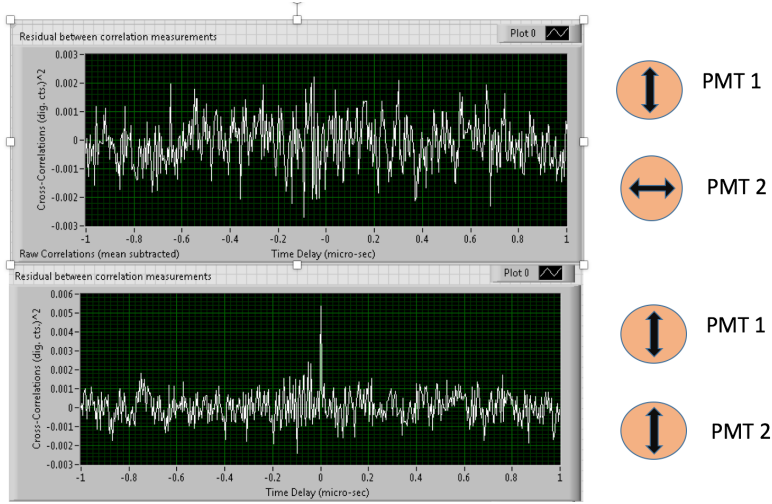

**Photon Bunching**

Figure 2 illustrates the 2-photon time coherence observed between two PMTS with zero spatial offset and different polarization. The measurements used 10-min observations of the Hg arc light source with a single 300 micron pinhole. The upper plot demonstrates the lack of time coherence with orthogonal light polarization. The lower plot demonstrates time bunching with parallel polarization (time delay 0). The time correlation only occurs with parallel light polarization. This verifies the HBT photon bunching, and excludes electrical noise origin.

*Figure 2: Laboratory measurement of time coherence from a Hg arc lamp source.. Upper plot: non-correlation with perpendicular polarizations. Lower plot: observed 2-photon correlation with parallel polarizer configuration. Horizontal axis: time delay between PMTs (μsec). Vertical axis: 2-photon cross correlation.*

The spatial coherence was measured using the linear actuator to scan a range of spatial offsets. The measurement used 10-min exposures of unpolarized Hg Arc lamp light from through the single 300 micron pinhole. Figure 3 compares the observed spatial coherence to a theoretical curve (including effects of the finite size of the PMT aperture). This demonstrates the ability of SII to accurately measure the diameter of a simulated star under similar coherence time and Signal/Noise ratios as a true main-sequence star.

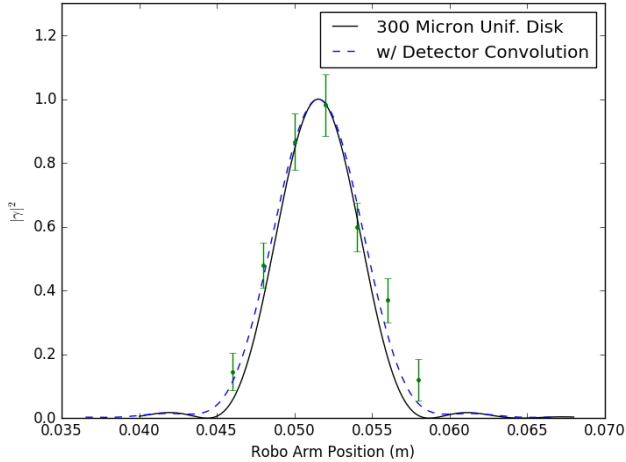

Figure 3. Observation of spatial correlation observed using automated linear scan. Horizontal axis: robot arm position (zero baseline is 0.052 m). Vertical axis: normalized square degre of coherence.

**Intensity Interferometry imaging** The SII laboratory uses different diameter hole combinations (200/300 micron) to simulate different single and binary star configurations. A typical measurement involves using the linear actuator to make evenly sampled measurements of the spatial correlation along linear baselines at different scan angles. This allows a periodically sampled measurement of $|\gamma(r)|^2$ in the Fourier reciporical image plane (Figure 4). Synthetic images are reconstructed from the interpolated Fourier reciporical image plane using a MIRA software analysis package [26] modified to compensate for the loss of phase information.

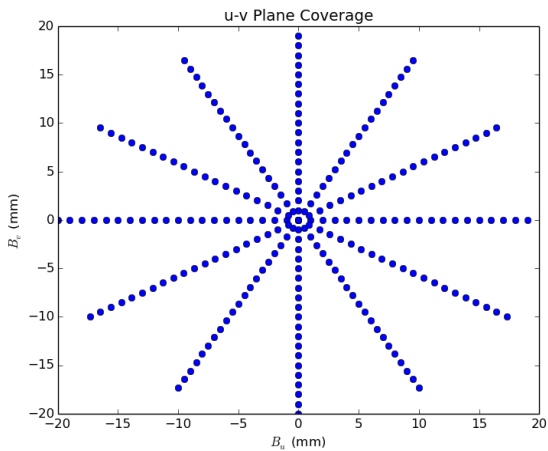

Figure 4: Scanning strategy for mapping the square degree of coherence across the Fourier image plane.

Populating the Fourier reciporical image plane with $|\gamma(r)|^2$ measurements in this manner requires about 100 measurements. We perform these SII imaging tests using the pseudo-random light source with a higher Q.E. SPAD detector which requires only 10 s to perform each measurement at a similar statistical significance as the PMT/ Hg arc lamp combination. The left panel of Figure 5 shows the MIRA reconstructed image of the 300 micron pinhole: the disk apears uniform and pinhole diameter is reconstructed with 5% accuracy. The right panel of Figure 5 demonstrates the reconstruction of a simulated binary pair: the relative diameters and pair separation are measured with 5% accuracy.

3. Strawman SII IACT implementations

A modular SII system for a large IACT array will use distributed, independent data acquisition systems at each telescope, with common synchronization of individual telescope data acquisition clocks and start/stop commands to < 1 nsec accuracy (Figure 6). This implementation must also provide a capability to transport the large data streams from each telescope to a common computation platform to calculate the cross-correlations between every photodetector data stream. The modularity of the proposed system allows additional telescopes to be incorporated into the array by simply connecting the new telescope to the central facility using fiber optic cables.

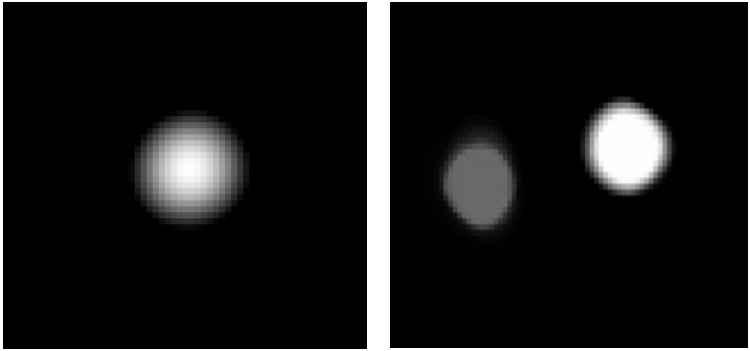

Figure 5: Reconstructed images of 300 micron diameter pinhole lab measurements (left panel) and simulated binary system (right panel). The images were reconstructed from the Fourier image plane using a modified MIRA image reconstruction package [24].

**Focal Plane Instrumentation** The typical SII focal plane instrumentation will integrate two high speed light sensors (PMTs or SiPM) with polarizing beamsplitter, narrowband optical filter (typical 5-10 nm bandpass) and collimator into a single, compact, integrated module (Figure 7). The SII pixel size is matched to the IACT focal plane p.s.f. The SII module is designed to be integrated into the IACT camera, or allow quick replacement of existing camera pixels. For example, an SII augmentation of VERITAS would use Hamamatsu R10560 PMTs, and could replace the three VERITAS pixels located near the optical axis. These pixels are easily accessible and can be exchanged with a SII pixel module in about 90 minutes (for a single telescope).

Each PMT is powered by an independent battery powered fiber-optic controlled HV supply. The HV setting and power can be controlled through pulse width modulation of the fiber optic transmitter signal. Each HV supply and battery is located nearby the SII pixel in each camera. The output of each PMT is amplified by a high speed (>200 MHz bandwidth) preamplifier which drives a double shielded (e.g. RG-223) coaxial cable. The coaxial cables transport the preamplifier signal to the SII DAQ system, located in/near the base of each telescope.

**Data Readout, Calibration and Multi-Telescope Synchronization** We expect an IACT implementation will use a commercially available system for initial implementation on small IACT arrays (e.g. VERITAS, Figure 6). The commercial DAQ system will eventually be replaced by a lower cost, custom single-board solution for larger IACT arrays (i.e. CTA). Both systems will share a common DAQ architecture with the Utah laboratory system described in Section 2: A high speed backplane (e.g. PXIe, >3 GB/sec) will link a high capacity (>10 TB) high speed (1-3 GB/sec) RAID disk to a 8-12 bit continuous digitizer controlled by an FPGA). The two PMT signals at each telecope are continuously digitized at > 250MHz by the FPGA with anti-aliasing filter. The FPGA digitization rate is matched to the on-axis isochronicity of the IACT optics (e.g 4 ns for VERITAS telescopes, <0.1 ns for CTA-SCT telescopes). The digitized data from each PMT is scaled and merged into a single continuous data stream by the FGPA and streamed to the RAID disk. The IACT optical calibration (flasher) system is used to equalize and calibrate the SII pixel gains across the full SII observatory.

The DAQ system uses phase locked loops to synchronize the PXIe backplane and FPGA clocks to a central GPS-locked timecode generator using White Rabbit timing modules (e.g. Seven Solutions WR-Switch/WR-LEN). The WR-LEN modules use singlemode fibers to transmit a phase-locked 10 MHz clock and 1 PPS with resolution of <200 psec RMS to each IACT. The WR-LEN modules can synchronize two clocks with 80km separation to better than 1 nsec RMS.

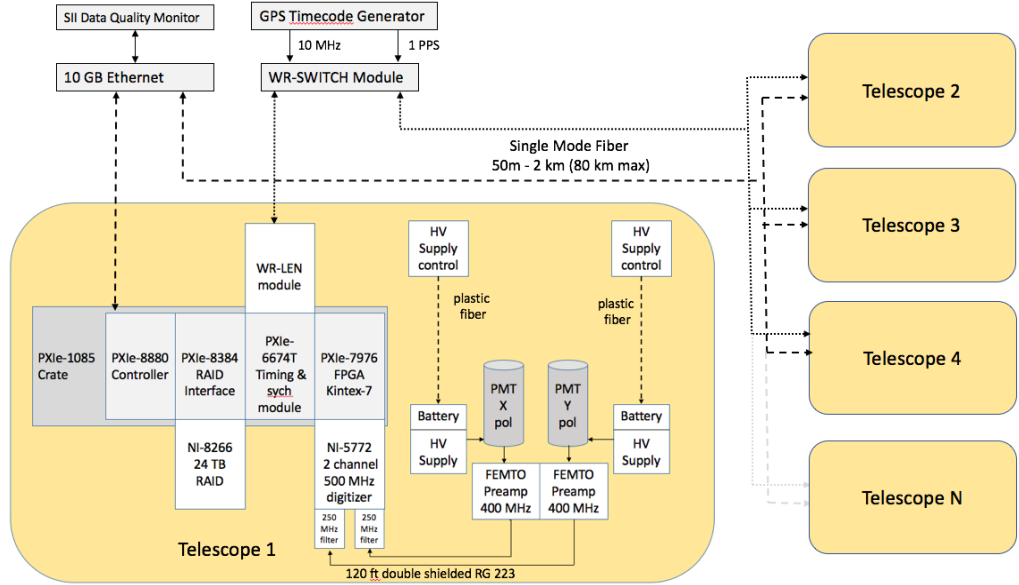

*Figure 6: Distributed data acquisition system for commercial SII instrumentation of an IACT Observatory*

A typical SII observation would begin by performing gain equalization/calibration of all SII pixels. The IACT observatory would then track the target star a SII observations are performed continuously performed for 4-6 hours about the source culmination (>10 TB/night/telescope). The projected distance between the telescope combinations will continuously change with changes in source elevation and azimuth. The measurements sample specific tracks of the square degree of coherence $|\gamma(r)|^2$ across the Fourier reciporical image plane. By breaking up the observations into discrete 30 minute intervals, the cross correlation between individual PMTs can be associated which short track segments in the Fourier reciporical image plane. The shorter data files also allow simplified, parallelized data transport and faster correlation. The start and stop of each 30 minute observation would be triggered at each telescope by the synchronized 1 PPS signal, allowing easy multi-telescope data merging. A 1 PPS GPS timestamp is also recorded with the event number and telescope pointing for each data stream, allowing reconstruction of the absolute UT time and sky location. An array-wide slow control system is used to automatically coordinate IACT pointing with the start/stop of individual observations runs and data transfer.

During the night, subsamples of the 30 minute observations are transported across a 10G fiber network to a central analysis computer and correlator. The analysis computer performs noise level calculations for each channel over short (1 minute) intervals, and online diagnostics provide information to the observer regarding data quality. Initial correlations between selected PMT pairs are performed in near-real time using an FPGA farm connected to high speed streaming SSD-RAID storage through a high speed (>3 GB) PXIe backplane. We recently have demonstrated the ability to perform real-time correlations between two channels using an FPGA. A single Virtex-5

FPGA can complete the correlation of a two-telescope correlation observation in about twice the obervation time. A farm of interconnected FPGAs can also perform 3-telescope (and higher order) correlations; these contain additional information (including phase closure information) useful for synthetic image reconstruction [27,28,29]. Source images are quickly reconstructed after the visibility is calculated across the Fourier reciporical image plane.

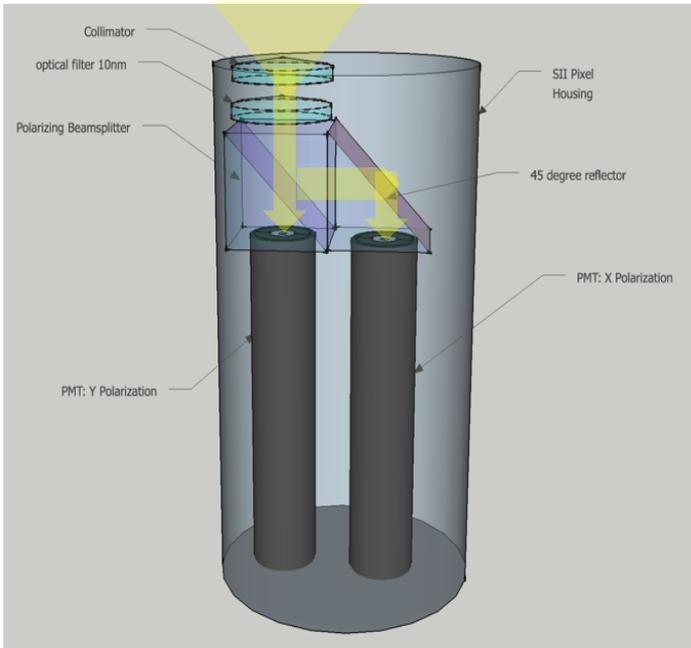

*Figure 7: Artist model of focal plane optical SII pixel. The self-contained pixel unit contains two independent PMTs for simultaneous measurement of two photon polarizations.*

### 4. Performance on Current and Future IACT arrays

For a large distributed array of 10-12 m diameter IACTs (i.e. VERITAS or CTA) it is estimated that stars as faint as $m_V = 6 - 7$ will be sufficiently bright for SII observation [8,30]. Selecting this magnitude range, approximately 1000 stars in the JMMC catalog [31] may be observable by VERITAS SII with an exposure of 1 hour [30], and 2500 may be detectable with a 10-hour exposure. With a 5-hour observation, we expect a zenith p.s.f < 0.7 mas for VERITAS SII imaging under realistic instrumental and observational conditions. Stars whose radii are easily resolvable with VERITAS SII include Rigel (2.4 mas, $m_v$=0.12), α Leo, γ Ori, βTau or ηUMa. Typically 0.5-2.0 observation hours are required for a 5σ determination of $|\gamma(r)|^2 \approx 0.5$ for these stars. It may also be possible to observe orbital modulation of the coherence in the binary systems Spica or Algol (1.5/2.2 mas semi-major axis respectively).

CTA SII imaging will have a factor of ~4-6 improvement in S/N over VERITAS imaging due to the larger number of telescopes and higher pixel quantum efficiency. The MST-SCT optical design [32,33] is isochronous on axis, with a time spread < 100 psec, allowing a factor of 10 increase in sampling speed. This improves the S/N ratio by an additional factor of 3. In addition, the 1-km telescope baselines in CTA allows the p.s.f to approach 100 $\mu$as resolution.